\documentclass{article}

\usepackage{arxiv}

\usepackage[utf8]{inputenc} 
\usepackage[T1]{fontenc}    
\usepackage{hyperref}       
\usepackage{url}            
\usepackage{booktabs}       
\usepackage{amsfonts}       
\usepackage{nicefrac}       
\usepackage{microtype}      
\usepackage{lipsum}
\usepackage{graphicx}
\usepackage{subcaption}
\usepackage{stfloats}
\usepackage{xcolor}

\usepackage[nolist,nohyperlinks]{acronym}
\usepackage{cite}
\acrodef{BIM}[BIM]{\emph{Building Information Modeling}}
\acrodef{CSR}[CSR]{\emph{Cholesteric Spherical Reflector}}
\acrodef{CLC}[CLC]{\emph{Cholesteric Liquid Crystal}}
\acrodef{AR}[AR]{\emph{Augmented Reality}}
\acrodef{AEC}[AEC]{\emph{Architecture Engineering and Construction}}
\acrodef{SKU}[SKU]{\emph{Stock Keeping Unit}}
\acrodef{CMYK}[CMYK]{\emph{Cyan Magenta Yellow Black}}
\acrodef{NFC}[NFC]{\emph{Near Field Communication}}
\acrodef{CBC}[CBC]{\emph{Construction Blockchain Consortium} } 
\acrodef{IFC}[IFC]{\emph{Industry Foundation Classes}}
\acrodef{SLAM}[SLAM]{\emph{Simultaneous Localization And Mapping}}

\title{Embedding Intelligence in Materials for Responsive Built Environment: A Topical Review on Liquid Crystal Elastomer Actuators and Sensors}

\author{
  Mathew Schwartz\thanks{http://smart-art.org)} \\
  New Jersey Institute of Technology\\
  College of Architecture and Design\\
  University Heights\\
  Newark, NJ, USA\\
  \texttt{cadop@umich.edu} \\
   \And
  Jan P.F. Lagerwall\thanks{www.lcsoftmatter.com)} \\
  University of Luxembourg\\
  Department of Physics \& Materials Science\\
  162a, avenue de la faiencerie\\
  1511 Luxembourg city, Luxembourg\\
  \texttt{jan.lagerwall@lcsoftmatter.com} \\
}

\begin{document}
\maketitle

\begin{abstract}
Liquid Crystal Elastomers (LCEs) are an exciting category of material that has tremendous application potential across a variety of fields, owing to their unique properties that enable both sensing and actuation. To some, LCEs are simply another type of Shape Memory Polymer, while to others they are an interesting on-going scientific experiment. In this visionary article, we bring an interdisciplinary discussion around creative and impactful ways that LCEs can be applied in the Built Environment to support kinematic and kinetic buildings and situational awareness. We focus particularly on the autonomy made possible by using LCEs, potentially removing needs for motors, wiring and tubing, and even enabling fully independent operation in response to natural environment variations, requiring no power sources. To illustrate the potential, we propose a number of concrete application scenarios where LCEs could offer innovative solutions to problems of great societal importance, such as autonomous active ventilation, heliotropic solar panel systems which can also remove snow or sand autonomously, and invisible coatings with strain mapping functionality, alerting residents in case of dangerous (static or dynamic) loads on roofs or windows, as well as assisting building safety inspection teams after earthquakes.

\end{abstract}


\section{Introduction}\label{preamble}
A lively development in today's materials science research is the creation of new 'smart' or 'intelligent' materials. This should not be confused with the quest for true artificial intelligence, but it generally refers to materials that respond to specific stimuli, such as mechanical strain or changes in temperature, humidity, lighting conditions etc. in such a way that their own appearance or shape changes. If the response has a beneficial impact, e.g., changing color, texture or shape to blend in with a new environment or prevent an undesired outcome, we can describe the material as adaptive. When we use adaptive materials to drive a change that is foundational to the performance or function of a technology or device-- in response to relevant stimuli-- we may say that the system has \textit{embedded intelligence}. One of the most impactful areas that such materials can be applied in is the Built Environment, enabling 'smart cities' with minimum impact on communication bandwidth and energy consumption \cite{Napolitano2021}. 

An important characteristic of embedded intelligence, defined in this way, is that the adaptive nature is inherent to the material, rather than arising from the interaction of multiple components with specific function, each made of passive and static materials like standard metals, plastics, wood or glass. In other words, while the same functionality may be realized by engineering traditional machines of varying complexity, we here mean with embedded intelligence that the responsiveness is provided by the material itself. This allows single elements in a construction to be their own actuator or sensor, while at the same time filling the basic passive function that they were designed for. A simple example is a hinge with embedded intelligence; while a classic hinge is passive, allowing motion but requiring an external engine for the motion to take place, the hinges we deal with in this paper move on their own, indeed \textit{driving} the motion of components attached to them. We thus consider not simply embedding intelligent machines in a system, but rather \textit{embedding intelligence of a system} via a material.

Among the most impactful consequences is the independence that arises, as fully implemented embedded intelligence requires no energy source---like a battery, solar cell or fuel tank---nor engines for doing mechanical work---like pumps or motors---nor wiring or tubing connecting components. As the actuation/sensing function is inherent to the material, relying on no other input than the stimulus triggering the response and the thermal energy driving it, adaptive materials enable unprecedented autonomy as well as ease in design. At the same time, however, the responsiveness cannot be turned off: with fully embedded intelligence, the material adapts to its environment whether we want it or not. For this reason, and for situations where we want to turn the response on at any time, it is sometimes motivated to develop the concept only partially, sacrificing complete autonomy for active control.

In the architecture, construction and design fields, adaptive materials capable of motion are often collectively known as \textit{Shape Memory Materials}, with Shape Memory Alloys perhaps being the most well known class\cite{persiani2020design,bengisu2018materials}. However, there are many types of material that classify as adaptive, with strongly varying characteristics, challenges and opportunities. In the materials science community, several visionary papers have recently appeared~\cite{McCracken2020,Xiao2020,Ko2017,Zeng2017a,Hines2017,Iamsaard2016,Yang2016,white2015programmable,Jiang2013Nanoscale}, presenting the potential of new and old adaptive materials as actuators and sensors, often focusing on applications in robotics. While they paint a bright and exciting picture, they also highlight the need for research to present adaptive materials solutions that compete favorably with established solutions, in particular those based on electronics and pneumatics. This can be challenging in state-of-the-art robotics, where rapid and high-powered actuation is often required. 

In this article we take a different approach, based in our belief that more prolific targets for nearby applications of adaptive materials may be those where there is no competition, because the new materials fill functions currently not existing--solving problems that presently have no satisfactory solution. A simple yet representative example is in architecture, where embedded intelligence through materials has been discussed through concepts such as incorporating Shape Memory Alloys into facades for opening walls when temperatures rise~\cite{doumpioti2010embedded}. Incorporation of embedded intelligence in the built environment offers many such targets, and these opportunities are the topic of this forward-looking article, aimed to stimulate creativity and discussion. We present several novel application scenarios, some of which may end up more challenging to realize than others, but all of them present highly stimulating opportunities for research and innovation. Our focus is particularly on truly autonomous applications, i.e., where the trigger for actuation appears naturally, e.g. through light variations across the day, humidity rising in the environment, or the mechanical strain on a window during a storm. 

We limit the discussion to one prolific class of adaptive material, namely \textit{Liquid Crystal Elastomers} (LCEs). LCEs are Shape Memory Elastomers and, as such, a subset of Shape Memory Polymers. They are highly responsive materials that can be designed to change shape, stiffness or color---often several of these characteristics together---in response to stimuli such as heat, light, exposure to water or mechanical strain. After introducing the working principle of LCEs and briefly surveying the different actuation modes they offer---all very attractive from an architecture, design and construction point of view---we provide a number of concrete examples of how appropriately designed and applied LCEs could embed intelligence into the built environment. Our examples illustrate how LCEs could support a transition towards more organic design solutions, where manmade artifacts work with, rather than against, daily and seasonal variations, giving them a lower energy consumption footprint. In addition, we show how LCEs might introduce highly valuable safety features into buildings occupied by people, potentially saving many lives in case of accidents or natural disasters. 

As we present LCEs in the context of innovation for the Built Environment, a brief discussion on contextualizing the materials may be beneficial to readers in that community. Our goal is to connect our own two disciplines---Design of the Built Environment and Materials Physics---to introduce to designers how LCEs work, and show to physicists what LCE functionality is the most useful in the Built Environment. We hope that our article can help to speed up the much anticipated transition from today's situation, with a plethora of fascinating LCE actuators and sensors that are custom-made in limited numbers in research labs, to a situation where LCEs can be practically applied to solve real-life problems, thanks to LCE components being mass produced and readily available for purchase from standard retailers.

\section{Principles of LCEs as actuators and sensors}

A simple way of illustrating the operation of LCE actuators is to consider them as pieces of rubber that stretch and relax on their own, driven by temperature changes or different light exposures. Indeed, the \textit{Elastomer} in LCE is another word for rubber, and on a fundamental structural level, LCEs and ordinary rubbers share three key components~\cite{Grosberg2011Giant}: (1) both are polymers, i.e., they are made up of very large molecules that can be described as chains of repeating units (monomers); (2) at operating temperature these molecules are in a highly mobile state which would normally correspond to the macroscopic state of a liquid, were it not for the significant fact that (3) these molecules are loosely connected into a volume-spanning network that prevents the system from flowing. In other words, elastomers can be considered liquids that are prevented from flowing by a 3D network connectivity of their molecules. The bonds that connect the molecular chains into the 3D network are called \textit{crosslinks} and their density is a key parameter for determining the mechanical properties of elastomers: high crosslink density gives a tough, stiff rubber, as in car tires, whereas low crosslink density gives a soft, extremely stretchable rubber, as in a rubber band. 

The crosslinked network has genuinely fascinating consequences, because it provides a direct link between the nanometer scale molecular conformation with the macroscopic shape of the entire object. In fact, when you pull at a rubber band, you are changing the shape of its molecules~\cite{Grosberg2011Giant}: you are forcing the molecular chains to stretch out along the direction you are pulling. In doing so, the molecules occupy less space in the plane perpendicular to the pulling direction, which is why the rubber band gets thinner perpendicular to the pulling direction. As soon as you stop pulling, however, the molecule chains are freed from their restraints and thermal energy (heat) will reorder them more or less randomly such that the macroscopic material shortens along the stretching direction and thickens in the perpendicular plane, thus reverting the rubber to its original state.

\begin{figure}
\begin{centering}
  \includegraphics[width=.6\textwidth]{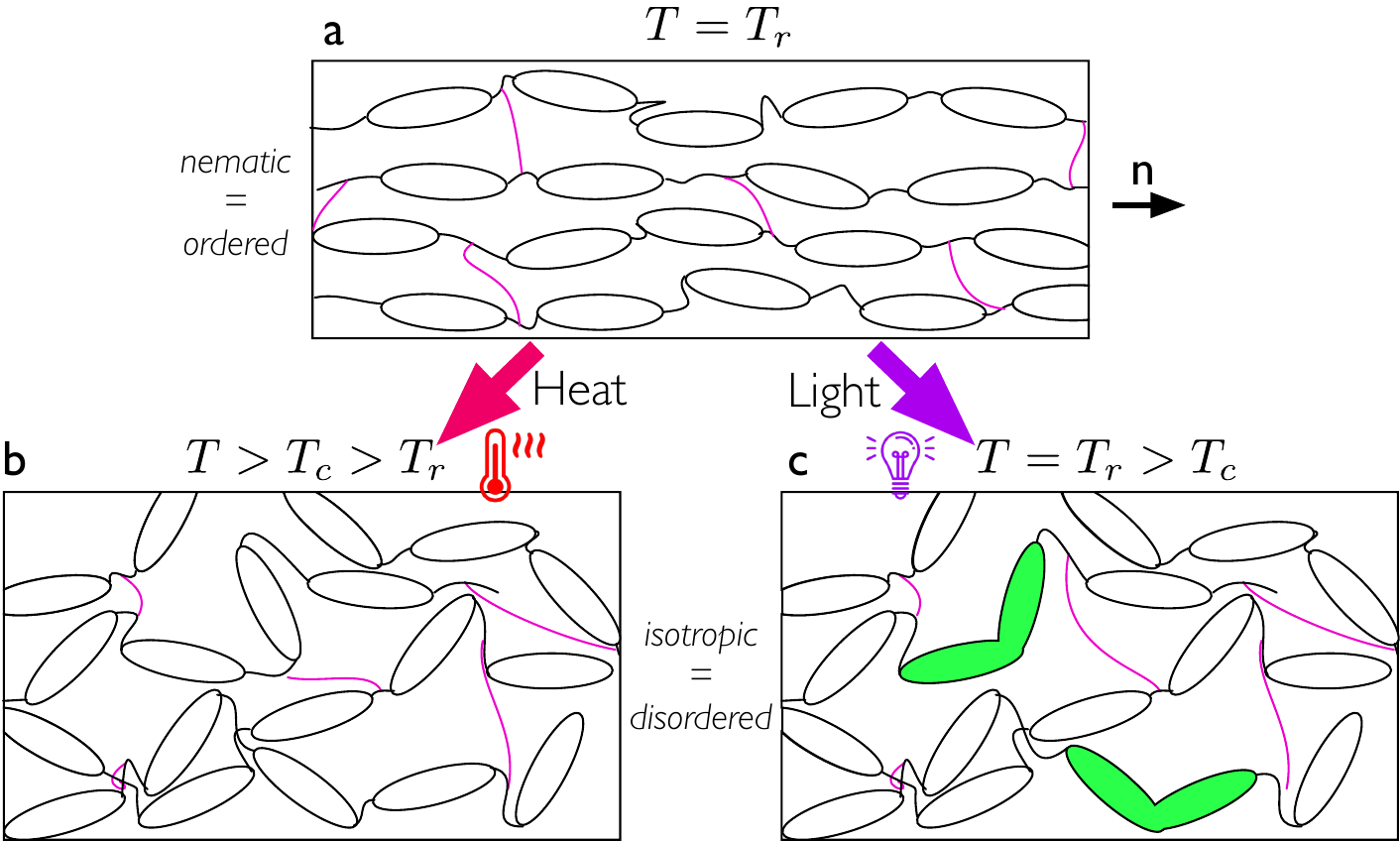}
  \caption{2D illustration of the working principle of an ordinary nematic LCE operated as actuator, triggered by heat and light, respectively. The ovals represent "mesogenic" segments, LC-promoting sections that prefer parallel alignment below $T_c$, in the ordered nematic phase. The green kinked shapes represent the order-disrupting illuminated state of light-responsive molecules such as azo-dyes, which reduce $T_c$ to below room temperature ($T_r$) and thus trigger actuation without heating. Whether triggered by light or heat, the actuated state is a disordered (isotropic) state, in which the elongation along the director (\textbf{n}) present in the nematic phase disappears. Because the system is crosslinked (pink bonds) into a 3D network, the molecular shape change induces a corresponding shape change of the macroscopic object, shortening along \textbf{n} and thicknening perpendicular to it.}\label{principle}
  \end{centering}
\end{figure}

The key difference between a regular rubber and an LCE is that the process just described can take place fully autonomously, within the LCE. The reason is that the molecules of an LCE tend to align along a specific direction at room temperature, as illustrated in \textbf{Figure~\ref{principle}}a. We call this direction, which can be defined macroscopically at the stage of crosslinking, the \textit{director} (abbreviated \textbf{n}). The liquid-like mobility combined with the orientational order makes the phase a \textit{Liquid Crystal} (LC) phase~\cite{lagerwall2012new}. The orientational order means that the molecular chains of an LCE, in contrast to an ordinary rubber, are stretched out along \textbf{n} \textit{without any external force applied}~\cite{ohm2010liquid,Fleischmann2013a}. However, if we heat beyond a critical temperature $T_c$, this spontaneous ordering is lost and the LCE switches to the behavior of an ordinary rubber that is spontaneously disordered. Because of the link between molecular order and macroscopic shape afforded by the 3D network of crosslinks, the consequence is that an LCE is elongated along \textbf{n} (and compact perpendicular to \textbf{n}) in the ground state, but it contracts along $\textbf{n}$ (and expands perpendicular to \textbf{n}) if heated above $T_c$, see Figure~\ref{principle}b. Not only does the LCE change its shape in response to the temperature change, but it can even exert a force when doing so, pulling or pushing at objects attached to the LCE. This is what makes it possible for LCEs to be actuators, or artificial muscles. 

Heating and cooling is the most basic way of triggering actuation, but regardless of how it is initiated, the heart of the process is the transition from the LC phase ground state to a disordered phase at temperatures above $T_c$. Depending on the intended application, the ground state $T_c$ can be tuned to a convenient range by adjusting the chemistry. Two commonly employed chemistries are those developed by Keller and co-workers \cite{thomsen2001liquid} and by Yakacki et al.\cite{Yakacki2015a}, typically yielding $T_c$ on the order of $100^\circ$C and $80^\circ$C, respectively (each category allows significant variability for fine-tuning $T_c$). Such relatively high transition temperatures are often desirable, since it allows quick relaxation by rapid cooling when the heating stimulus is removed, but LCEs can be tuned to actuate at much lower temperature, for instance near body temperature as in the work of Shaha et al.\cite{Shaha2021} Importantly, we can also achieve the phase change without heating or cooling, by changing $T_c$ instead (Figure~\ref{principle}c). By incorporating light-responsive molecules which change their shape between rod-shaped and kinked when exposed to light of different wavelengths, we can tune $T_c$ back and forth below and above room temperature, by exposing with ultraviolet (UV) and visible light, respectively \cite{Yamada2008}. The LCE can now be actuated by light at constant temperature. In the cited example by Yamada et al., the LCE was actuated in both directions at room temperature, using UV light with wavelength $\lambda=366$~nm and intensity 240~mWcm$^{-2}$, and visible light with $\lambda>500$~nm and intensity 120~mWcm$^{-2}$. We can also combine light- and temperature-driven actuation by incorporating dyes in the LCE, which absorb light but immediately release the energy as heat: the LCE itself heats up internally due to the light exposure, yet the surrounding does not need to be heated. Following this strategy, Kohlmeyer and Chen induced actuation of an LCE doped with an IR-dye by irradiating at wavelength $\lambda=980$~nm and intensity 5.7~mWmm$^{-2}$ ~\cite{Kohlmeyer2013}. Interestingly, the same intensity of IR light at $\lambda=1342$~nm induced no actuation, demonstrating the possibility to fine-tune the responsiveness to stimuli only within a specific wavelength region. Finally, recent advances in the chemistry of LCEs has also allowed humidity~\cite{deHaan2014Humidity-responsive} and electric field-driven~\cite{liu2017protruding} actuation.

While the first LCEs were limited in terms of the freedom to choose the ground state shape and actuation mode. The latter were defined by a uniform orientation of \textbf{n} throughout the sample that was obtained by stretching it unidirectionally during production~\cite{kupfer1991nematic}, a major breakthrough were the demonstrations of controlled bending, allowing the LCE to function as an active hinge. This was first realized by irradiating light-responsive LCEs with collimated light \cite{Yu2003b,Camacho-Lopez2004}, later by deforming the ground state director field, \textbf{n}(\textbf{r}) (where \textbf{r} is the space coordinate), into a bend~\cite{van2009printed}. This last step was seminal, as it inspired a creative explosion of innovative approaches to pattern complex \textbf{n}(\textbf{r}) into LCEs over the last decade \cite{deHaan2012,McConney2013,Ware2015,Iamsaard2014,deHaan2014}. These references are far from exhaustive, but they contain some of the most striking examples.

Moreover, also the ground state shape is no longer restricted to the flat monolithic sheet of the original unidirectional stretching approach, but now LCEs can be realized as self-closing ribbons~\cite{Yamada2008,Geng2013}, spheroids~\cite{Ohm2011b,marshall2013anisotropic}, Janus particles~\cite{Hessberger2016a}, shells~\cite{fleischmann2012one,jampani2018micrometer,jampani2019liquid}, fibers~\cite{Ohm2011}, tubes~\cite{Braun2018}--indeed in any conceivable shape. When actuated; ribbons can function as motors that drive axle rotation~\cite{Yamada2008,Geng2013}, shells can function as micropumps~\cite{fleischmann2012one} and tubes, if they are fully developed with the right \textbf{n}(\textbf{r}), should function as a peristaltic pump~\cite{Selinger2008}. Among the most inspiring demonstrations is that of the Verduzco group, which defined complex 3D ground state shapes such as doll's faces and flowers by spanning the LCE precursor over the corresponding 3D molds, obtaining reversible actuation from the arbitrary shape into a flat sheet by heating above $T_c$~\cite{Barnes2019Direct}. As the procedure is simple and the LCEs are quite large, this truly holds promise for applications. Combining the freedom of determining the ground state shape with the freedom of tailoring \textbf{n}(\textbf{r}) into complex patterns, highly complex actuation modes can be programmed into LCEs. Adding the different options for triggering the shape change, LCEs arise as incredibly versatile and advanced actuators. By realizing components in LCEs or LCE composites, we can thus truly speak of embedded intelligence for these components. To place LCE actuators in the context of some commonly used comparable actuator solutions, we have compiled the key characteristics of each approach in Table~\ref{tab:sma-comparison}. Since this article focuses on LCEs, which is also our main expertise, we make no claims that the data are exhaustive; the numbers we state are typical numbers that can be found in articles and on-line media, representing our current best understanding. Exceptions may very well occur.

\begin{table}[!h]
\small
\caption{Comparison of LCE actuators with major competing shape memory actuators. Material abbreviations are: Shape memory polymers (SMP), Shape memory alloys (SMA), Dielectric elastomers (DE).}
\label{tab:sma-comparison}
\begin{tabular}
{p{1.8cm}p{1.4cm}p{1.4cm}p{1.7cm}p{1.4cm}p{1.9cm}p{1.9cm}}
                                & \textbf{SMP} & \textbf{SMA} & \textbf{DE} & \textbf{Hydrogels} & \textbf{Soft pneumatics} & \textbf{LCEs} \\ \hline
\multicolumn{1}{l|}{Actuation principle} & Glass transition & Crystal phase transition & Electrostatic attraction & Switchable solubility & Inflation \& deformation of rubber channels & Liquid crystal phase transition\\
\multicolumn{1}{l|}{Stimuli} & Temperature, light, chemical & Temperature & Voltage & Temperature & Pressurized air & Temperature, light, chemical, voltage\\
\multicolumn{1}{l|}{Easy to tune response threshold} & Yes & No & No & No & Yes & Yes\\
\multicolumn{1}{l|}{Bidirectional \& reversible actuation} & No & Yes & Yes & Yes & Yes & Yes\\
\multicolumn{1}{l|}{Actuation Force}        &  Medium & Medium & Strong & Weak & Strong & Weak\\     
\multicolumn{1}{l|}{Max strain} & $\sim800$\% & $\sim8$\% & $\sim100$\% & $\sim1000$\% & $\sim50$\% & 800\%\\     
\multicolumn{1}{l|}{Actuation speed} & Slow & Fast & Fast & Slow & Fast & Slow\\     
\multicolumn{1}{l|}{Mechanical character}  & Soft & Hard & Soft & Very soft & Soft & Soft\\
\multicolumn{1}{l|}{Energy source} & Thermal energy &  Thermal energy & High-voltage power supply & Thermal energy & Air compressor & Thermal energy \\
\multicolumn{1}{l|}{Other requirements} & None & None & Electrical wiring & Water & Tubing & None \\
\multicolumn{1}{l|}{Density} & Medium & High & Medium & Medium & Low & Medium \\     
\multicolumn{1}{l|}{Form factor}  & 1, 2 or 3D & 1D wire & 2D sheet & 1, 2 or 3D &  1, 2 or 3D &  1, 2 or 3D\\ 
\multicolumn{1}{l|}{Can be transparent}  & Yes & No & No & Yes & No & Yes\\ 
\multicolumn{1}{l|}{Current cost/kg}  & Low & High & Medium & Low & Low & High\\
\multicolumn{1}{l|}{Attainable cost/kg}  & Low & High & Medium & Low & Low & Low                  

\end{tabular}

\end{table}

Like many actuators, LCEs can also be operated in reverse, giving a signal in response to mechanical strain-- a sensor. The ideal solution for this is the cholesteric LCE (CLCE), which exhibits a helical modulation of \textbf{n} with period (called \textit{pitch}, \textit{p}) on the scale of light wavelengths, giving rise to so-called selective reflection of light with wavelength $\lambda\approx p$~\cite{lagerwall2012new}. By varying the chemical composition, $p$, and thus $\lambda$ can be varied at will, allowing us to tune this "structural color" from the infrared (IR), across the full visible band, and into the UV. 

As the helix is linked to the macroscopic object shape via the 3D network of crosslinks, an extension of a CLCE sheet or fiber perpendicular to the helix will lead to a compression of the helix, thus a reduction of $p$ and $\lambda$, with a consequent blue-shift of the reflection color~\cite{Finkelmann2001,Cicuta2004,Kizhakidathazhath2020Facile,Geng2022}. This means that an originally red-reflecting CLCE can give us continuous mechanical strain sensing (it works equally well for tensile and compressive strain~\cite{Kizhakidathazhath2020Facile}) with immediate read-out visible to the naked eye, in form of a color change from red to orange, yellow, green, blue and finally violet. To formulate this quantitatively, we demonstrated a shift of the reflection wavelength $\Delta \lambda = 145$~nm from red ($\lambda=625$~nm) to blue ($\lambda=480$~nm) upon 120\% tensile strain in the plane of a 2D CLCE sheet~\cite{Kizhakidathazhath2020Facile}, and a shift $\Delta \lambda = 155$~nm from red ($\lambda=615$~nm) to blue ($\lambda=460$~nm) upon 200\% extensile strain of a 1D CLCE fiber~\cite{Geng2022}. The response is also highly localized, i.e., if the strain varies across the CLCE, it shows different color in different points, enabling strain mapping with very high resolution~\cite{Kizhakidathazhath2020Facile,Geng2022,Martinez2020Reconfigurable} .
 
An interesting possibility is to prepare the CLCE sheet with $p$ long enough to reflect IR, but not visible, light. The CLCE thus looks colorless transparent in the ground state. But if it is stretched perpendicular to the helix, $p$ and $\lambda$ reduce, and if the original helix is properly tuned, a red color appears. De Haan and co-workers patterned such a CLCE within a passive transparent rubber sheet such that a red warning sign became visible beyond a critical strain threshold~\cite{Zhang2020}. Clearly, such CLCE strain sensors can be enormously useful tools in the built environment, as we will exemplify below. 

\section{Examples of opportunities of LCE actuators and sensors embedding intellingence into the built environment}

In the Architecture and Design fields, leveraging materials for use in building function and even user interaction is of great interest. While so many modern construction materials are created around standards to reduce both risk (standardized constraints) and cost (standardized shape in manufacturing), a desire for customization continues to drive innovation. While mass customization has gained traction in architecture research, this area is dominated by CNC (Computer Numerical Control) and robotic manufacturing techniques. Still, there has been an area, which continues to grow, of both industrial and academic works where customization and environment-specific interactive buildings are enabled through materials themselves. Most recently, Persiani published a detailed book discussing numerous actuating materials in the context of kinetic building design, referring to materials that enable kinetic responses to stimuli as \textit{Autoreaction}~\cite{persiani2020design}. This follows on the book by Bengisu and Ferrara, published only two years earlier, detailing multiple case-studies and uses of \textit{Materials That Move}~\cite{bengisu2018materials}. 

\begin{figure}[!htbp]
\centering
\includegraphics[page=5, width=.9\textwidth]{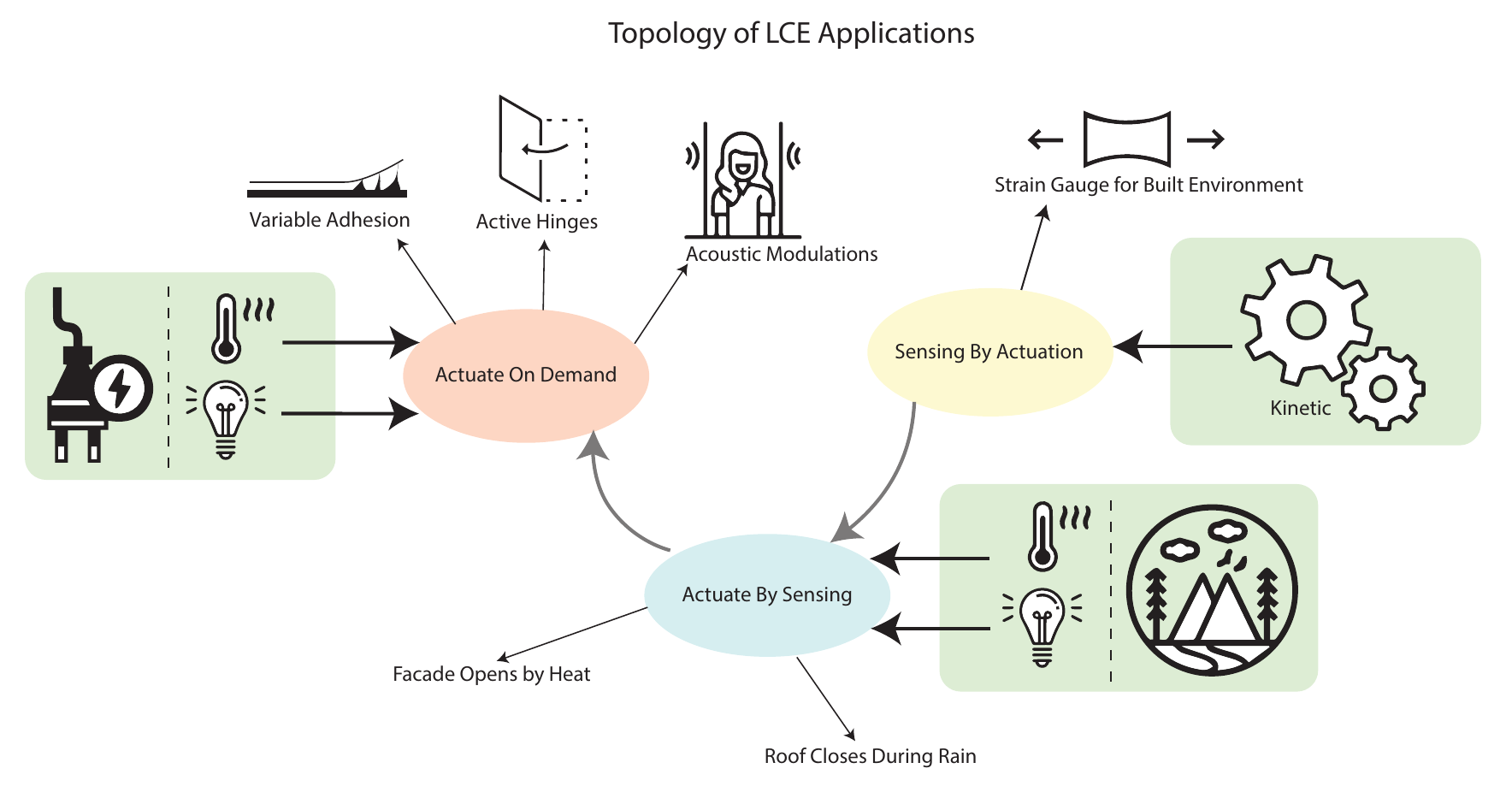}
\caption{Topology of the three themes of LCE interactions we discuss. The arrows connecting the themes illustrate the relationship between the types of inputs and outputs, from simple sensing, to sensing leading to actuation, to actuation arising from sensing an artificially created stimulus.}
\label{fig:topology} 
\end{figure}

In the vein of these books, we here give a few concrete application examples of embedded intelligence in the built environment, with the specific focus on the opportunities offered by LCEs (\textbf{Figure~\ref{fig:topology}}). We first consider them operating in tandem with additional devices that allow a user to freely choose when the actuation should take place, continuing with some scenarios where we can safely do without such devices, thereby reaching full embedded intelligence and truly autonomous operation. Finally, we go beyond the core concept of 'materials that move' and discuss the valuable potential of CLCE sensors triggered by motion, delivering a color response that can be life saving for people inhabiting a building.

\subsection{Actuation on Demand}

In this theme, we consider the situation where a user should be able to turn on or off actuation at will. From a materials design perspective, it is then an inherent requirement that the LCE does \textit{not} respond to the natural changes in environment, in terms of temperature, light, humidity or other factors. The chemistry must be tailored for a $T_c$ much higher than any expected ambient temperature, for instance using the Keller \cite{thomsen2001liquid} or Yakacki \cite{Yakacki2015a} approaches (see above), and if light-responsive molecules are incorporated, these should not be sensitive to sunlight or normal artificial lighting, but rather respond to UV light of rather low wavelength and/or high (local) intensity. The LCE must be complemented by artificial sources of actuation stimulus, e.g., a heater or light source which the user controls to actuate or relax the LCE. Of course, this solution consumes energy and will require some kind of power/energy supply, hence we have---by definition---sacrificed autonomy for control.  We refer to this way of using LCEs in a controlled way entirely independent of the environment as \textit{Actuation on Demand}. 

It is our impression that the most common application scenarios considered so far by the LCE research community belong to the Actuation on Demand category, often in the context of soft robotics. While the term robotics is broad, work relating to actuating LCEs often revolves around some sort of exo-skeleton or locomotive device. Importantly, many of the design paradigms used for mobile robotics can be translated to the situation of immobile built environment, solving important problems that have not been so much in focus for the LCE community. We see buildings not as monolithic static structures, but instead as objects that should---but are not always able to---conform to the occupants within them. Our static environments are not always the result of intention, but rather a limitation of the materials and technologies available to designers. There have been various projects from architects and designers that explore actuation on demand for replacing static with kinetic components. Here we will look at two specific illustrative examples, one initiated in the architect and design community, the other in the LCE community. 

\subsubsection{Modulating acoustics}
Sound is an integral part of how many people experience an environment, from hearing someone speaking during a lecture, to enjoying the art of musicians performing in a concert hall, to reducing noise in a crowded station. The two most common ways designers modify the sound of a space are to tune (1) the sound absorption capacity of materials used, and (2) the geometry to guide and scatter. In the approach of \cite{thun2012soundspheres}, a ceiling-type system of panels containing perforations are aggregated in an origami-like structure that is actuated through a mechanical system to open and close, thereby changing the acoustics. The system relies on electrical stepper motors, hence also continuous wiring from the power source to each individual element.

An extension of this concept using LCEs, replacing the stepper motors with (perfectly silent) light-responsive LCE hinges \cite{Kim2019Intelligently}, might allow an elimination of the wiring, as the light signal for triggering actuation (which can be invisible) can be sent through the air. If electrical wiring is acceptable, this can be used to drive heaters for actuating LCE hinges without the use of light \cite{Roach2018}, still without any noise generation. A significant challenge is that the panels may be of significant size and weight, and many LCE hinges would need to be applied in tandem to deliver sufficient force. Alternatively, the panels could be fitted with passive hinges with the opposite side suspended by LCE fibers, which contract upon actuation and expand upon relaxation. Roach et al. demonstrated production of meter-long continuous LCE fibers that can be knit, sewn or woven, featuring 51\% actuation strain \cite{Roach2019}. Lin et al. demonstrated continuous spinning of LCE fibers which actuate with a 40--60\% length reduction within 1~second \cite{Lin2021a}. Significant gains can be achieved by layering LCE sheets\cite{Guin2018} or bundling LCE fibers \cite{He2021a}, but more important may be to design the overall structure with counterbalances, such that the force needed from each LCE is minimized. Lifting large heavy plates against gravity with LCE hinges or fibers is most likely unrealistic. At least it would require excessive amounts of LCE material.

This application opportunity also puts the spotlight on an area of LCE research that is yet under-explored: bistability. Normally, LCEs relax as soon as the stimulus is removed, which means that the switch to a different type of acoustics would require continuous irradiation or heating of the LCE hinges. It would obviously be highly desirable to be able to lock the new shape into place, allowing the light or heat source to be turned off. Some success in promoting bistability was demonstrated using LCE-wire composites \cite{Huang2010Mechanical} but more research is required. In fact, this type of application would benefit the most from \textit{multistable} actuators, which could be locked into a multitude of different shapes. This might be achievable by using LCEs that are in a glassy, stiff state at room temperature, and combine actuation via local temperature change or light-responsive tuning of $T_c$ with heating or cooling above or below the glass transition $T_g$. An ideal solution could be to incorporate carbon nanotubes or IR-dyes \cite{Kohlmeyer2013,Kim2019Intelligently} as well as UV-responsive azobenzene moieties \cite{Yamada2008} in the LCE. The former allow very efficient localized heating by irradiation of the LCE by light. By using infrared light, the irradiation is invisible to humans while being very efficient in raising the temperature above a $T_g$ that has been tuned to be well above room temperature, yet well below $T_c$ in the absence of UV-light. The LCE could thus be switched from stiff to responsive by turning on IR radiation, and by then exposing with UV light, $T_c$ could be decreased to below the temperature achieved by the IR-heating, thereby inducing actuation. After actuation has been accomplished, the IR source is turned off before the UV source, rendering the LCE glassy and thus stiff again, but now in the actuated state. The UV source can then be turned off without the LCE relaxing. Once relaxation is desired, the IR source is turned on, heating the LCE above $T_g$ and allowing relaxation into the ground state, on demand, without continuous energy consumption in between.

\begin{figure}[!htbp]
\centering
\includegraphics[page=3, width=.8\textwidth]{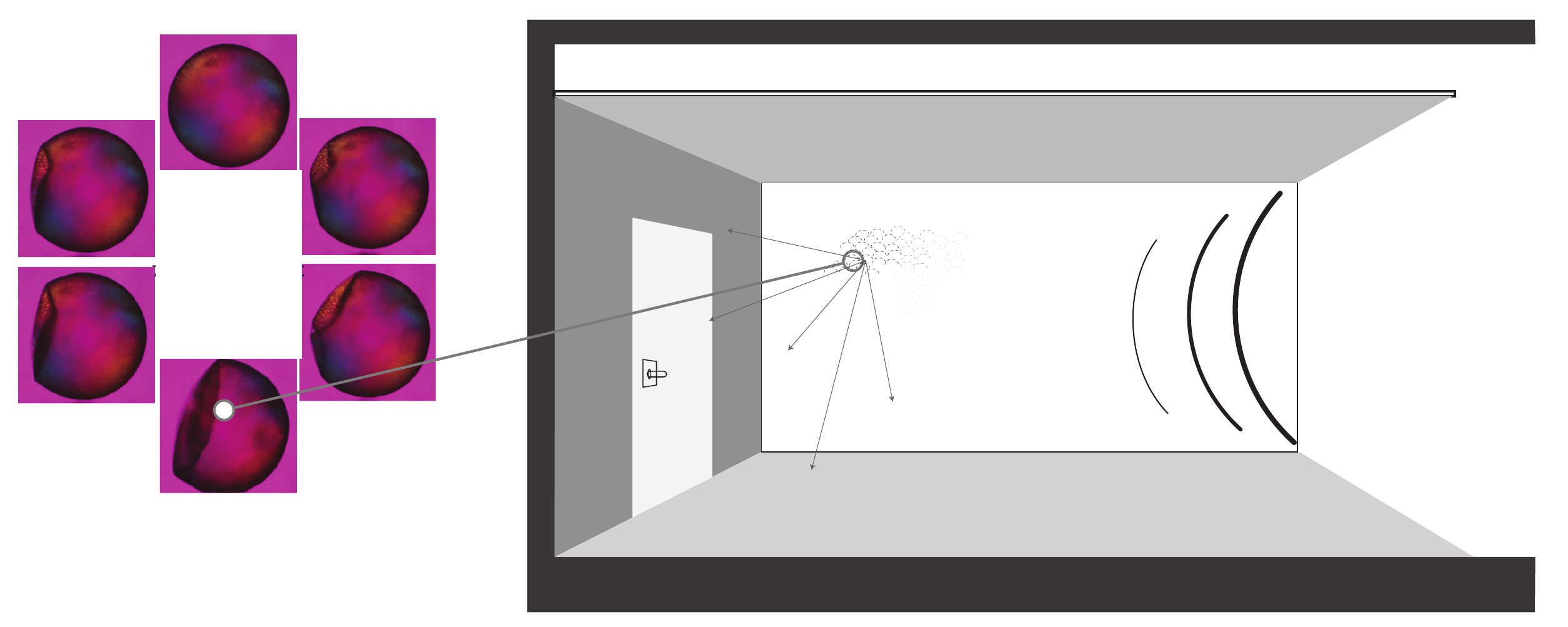}
\caption{ A wall is composed of LCE coatings that change form at a smaller scale to modify acoustics. The arrows represent sound scattering from an exterior source due to the geometric changes of the wall surface. The inset on the left shows an LCE shell exhibiting reversible buckling upon heating and cooling (from Jampani et al. \cite{jampani2018micrometer} CC BY-NC-ND 4.0 license), as an example of how a 'bubbly' texture could be modulated on demand using LCEs for tunable acoustics.}
\label{fig:acousticConcept} 
\end{figure}

In a different approach, Decker uses the concept of soft robotics in architecture to modulate the walls of a chamber by expanding a rubber sheet into a pre-molded bubble-like topography using pneumatics~\cite{decker2015soft}. Again, LCEs might elevate this solution for controlling acoustic properties by removing the need for connectivity---in this case via tubing rather than electrical wiring---as an LCE can be actuated on demand through heat (\textbf{Figure~\ref{fig:acousticConcept}}) and light. Following the approach of Barnes and Verduzco~\cite{Barnes2019Direct}, the bubble-like topography could be imprinted as the ground state, actuation yielding an increasingly flat surface. Another interesting approach would be to decorate a surface with LCE shells, spheres of diameter 0.1--10~mm which can be reversibly switched between taught, smooth and irregularly buckled, soft states~\cite{jampani2019liquid,jampani2018micrometer}. The actuation/relaxation may affect both the absorption and scattering of sound waves, the effect depending on the scale of the buckles as well as of the overall spheres. To control such LCE-based sound modulation, light panels could back-illuminate the LCE surface, or ceiling track-lighting could illuminate the LCE-covered surfaces in different patterns/intensities/configurations to provide unique acoustic modifications to the environment. Depending on the chemical design of the LCE, the light could be visible, for effect, or invisible, for discreteness.

\subsubsection{Modulating adhesion}
Photovoltaics (PV) technology, or solar energy, has enormous potential to solve our future energy needs, but before it can reach that goal, further advances are required both in terms of the basic PV performance and in terms of the practical reliability of PV installations. In this article, we consider two examples of how LCEs can help in the second respect--starting with the importance of keeping solar panels clean. If a solar panel is covered by debris, it will not deliver much electrical energy no matter how performing the actual PV module is~\cite{sayyah2014energy}. Unfortunately, some of the sunniest areas in the world, thus where PV is the most useful, are also regions where sand and dust are ubiquitous, hence efficient PV installations must be able to remove layers of dry particles covering the solar panel. Also in colder regions of the world, PV is considered a viable energy alternative, and here the corresponding challenge is to remove snow from the solar panels, which prevents the sunlight from reaching the active PV elements to generate electricity.  

Liu and co-workers have taken on this challenge in their research, developing films of liquid crystal networks (LCNs), which can be viewed as very densely crosslinked LCEs, in which the surface naturally develops a fingerprint-like or randomly varied topography. When these films are actuated, the troughs and valleys exchange location, and the amplitude of the topographic modulation can be increased. Importantly, the team recently reached a significant break-through in the LCN chemistry, allowing the actuation to be directly electrically triggered~\cite{liu2017protruding, Feng2020Static}, without intermediate light or heat stimuli. As a result, they could demonstrate release-on-demand of sand covering an inclined glass surface~\cite{Feng2018Oscillating}. By applying a voltage over the film, the dynamic modulation of its surface reduced the adhesion of the sand to such an extent that gravity pulled the sand grains down, leaving a surface free of debris. While much research still remains to be done, this is a highly inspiring concept, not least as the reliance on electrical power is a minor issue when dealing with PV cells, which deliver orders of magnitude greater electrical power than needed to activate the LCN. Different types of sand and dust may adhere to different extents, hence the concept may need to be boosted by varying the periodicity and amplitude of the topography modulation and also introduce specific chemical functionality in the coating. 

An alternative procedure for dynamic adhesion modulation with LCEs, inspired by the gecko's ability to switch between adhesion and release via the hairy surface topography of its attachment pads, is to realize an LCE film in the form of arrays of pillars with switchable height~\cite{Cui2012Bioinspired,Wu2013c,Wu2013b}. Combining with rigid rails that align with the pillar array in its relaxed state but protrude above the pillars when they are actuated, thus terminating contact between pillars and external objects, Cui et al. could switch between adhesive and non-adhesive states of their substrates. Of particular interest for snow removal may be the dynamic tuning of wettability achieved by Wu et al.~\cite{Wu2013c,Wu2013b}. Here no rigid array was needed but the effect was realized by tuning the aspect ratio of the pillars, thereby switching from Cassie to Wenzel style wetting. Such a surface may be able to act on a liquid water layer between the solar panel and the snow, promoting the snow to slide down. Interestingly, Farre-Kaga et al. recently demonstrated that LCEs are exceptional pressure-sensitive adhesives while in the nematic phase, but the enhanced performance is lost upon heating to the isotropic state \cite{Farre-Kaga2022}, hence there is much room for future applications of LCEs as dynamic adhesives. 

Of course, the best way to deal with debris covering solar panels is if the source of the debris can be avoided in the first place. Below we will see how LCEs, in conjunction with a new type of miniaturized PV elements, may also be useful in this way.

\subsection{Actuation by Sensing}
In many areas of the world, buildings are a major contributor to greenhouse gases through their massive energy requirements. While the structural aspect of a building is typically a one-time cost (i.e., concrete is poured once), the ongoing regulation that balances the variation of nature with the human expectation of consistency is a major challenge.  Simply put, we spend an incredible amount of energy making buildings warm when nature is cold, light when it is dark, and dark when it is too light. 

Here we consider application scenarios for LCEs where they assist buildings to provide a consistent environment for the occupants, with minimum energy requirements--by actuating fully autonomously in response to the natural environmental variations. This means that we often need to work with a rather low $T_c$ in order to reach actuation without artificial heating. A weakness of LCEs operating at low $T_c$ is the slow relaxation, arising from the time needed to cool down below $T_c$ if this is near room temperature. Fortunately, compared to applications in robotics, the low speeds of change in the scenarios considered here are more forgiving in this respect, as traditional methods such as gratings~\cite{ZAKIRULLIN2022109258} are static, rendering this weakness acceptable. 

\subsubsection{Regulating facades and roofs in response to periodic temperature and light variations}

Consider the basic scenario of balancing natural light and heat radiation input with the active temperature control of an air conditioning system, consuming energy for heating in winter and for cooling in summer. While in warmer months we could rely on coated glass that blocks infrared light from the outside, we would hope to increase the infrared light input from the outside during colder months. On a shorter time scale, such dynamic facade modulation would be valuable even as it is cooler in the mornings and evenings. An exciting solution to this challenge is in phase-change materials (PCMs), as simple as wax, that absorb heat during melting upon heating above a threshold temperature, and release heat during solidification if cooled below this temperature. In the built environment literature there have been numerous studies using PCMs for thermal regulation with a variety of materials and methods~\cite{WANG2022109436,GONCALVES2021108281,GUO2022109318,WANG2022109155}. 

The phase change in question is also coupled to a very significant specific volume change, standard paraffin wax expanding on the order of 20\% in volume upon melting. By preparing spherical shells of cholesteric liquid crystal with a paraffin wax core, and polymerizing the shell into a CLCE that is close to index matched to the core, the volume expansion would lead to a shell thinning with consequent reduction of $p$ and a consequent blue-shift of the retroreflection color~\cite{Kizhakidathazhath2020Facile}, which is omnidirectional in the shell~\cite{schwartz2018cholesteric}. By tuning the initial pitch to values long enough that the most important heat radiation is let through in the cold state, moving into the heat blocking part of the spectrum upon compression, wax-filled CLCE shells might thus be able to couple the classic latent heat-based temperature regulation well known from phase-change materials with variations in IR transmission. An advantage of using shells is that the angle dependence of the reflected wavelength range is greatly reduced compared to flat cholesteric films \cite{schwartz2018cholesteric}, which is important as the sun changes position over the day and over the season.

While the realization of wax-filled shells with tuned and index matched CLCE is a challenge to be met, light-driven autonomous irises have already been realized from LCEs~\cite{Zeng2017Self-Regulating}. Here a circular LCE sheet is prepared with \textbf{n}(\textbf{r}) patterned into a bend from top to bottom, and radial cuts are made towards the center. The result is that the LCE sections curl upwards under low illumination intensity, exposing a central hole through which light can enter. When irradiated with sufficient UV light, however, the sections flatten out towards the center, thereby blocking the light path. While this iris thus expands in the third dimension during actuation, a strictly 2D actuation-on-demand version of an LCE iris was demonstrated by the Zentel group~\cite{Schuhladen2014}. Here a radial array of electrical heaters integrated within the LCE sheat increased the size of the central hole in the iris without any bending. 

\begin{figure}[!htbp]
\centering
    \begin{subfigure}[b]{0.48\textwidth}
    	\includegraphics[page=6,width=1\textwidth]{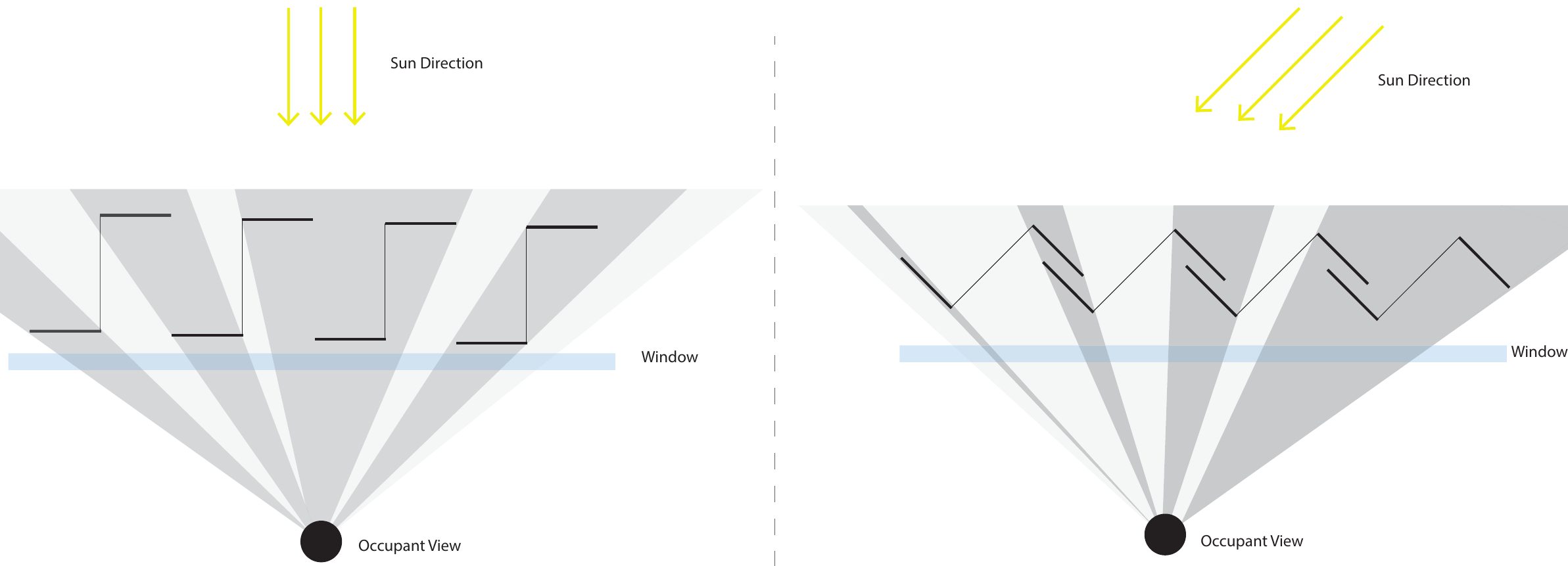}
    	\caption{}
    	\label{subfig:left}
	\end{subfigure}
    \begin{subfigure}[b]{0.48\textwidth}
    	\includegraphics[page=7,width=1\textwidth]{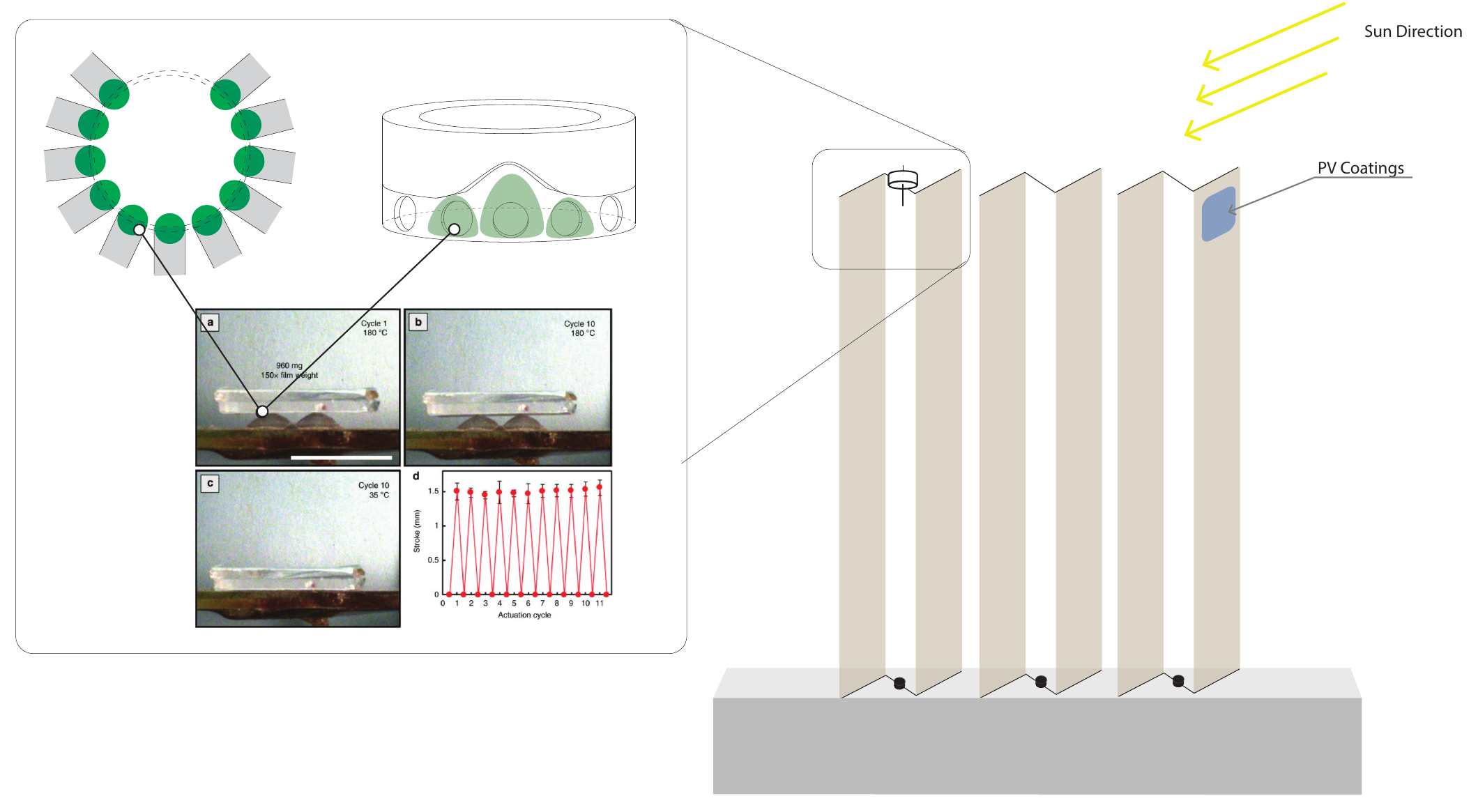}
    	\caption{}
    	\label{subfig:right}
	\end{subfigure}
    \caption{(Fig\ref{subfig:left}) Sun-tracking louver in plan (top view), where the rotating blades block direct sunlight but allow indirect light and a view to the exterior. (Fig\ref{subfig:right}) Concept drawings of how the heliotropic function could be realized using LCE sheets that protrude as pillars when actuated. As an example of LCEs with significant lifting power we show the patterned and layered LCE sheets realized by the White team \cite{Guin2018} (CC BY 4.0 license). If the louver surfaces are covered with micro-PV panels, the system will automatically direct them towards the sun throughout the day. If each PV panel is attached via a light-activated LCE hinge, it will show its active face only during the light part of the day, while it is folded inwards against the louver, protected agains sand, snow and dust, during the night.} 
    \label{fig:louverConcept} 
\end{figure}

An even more accessible means of achieving autonomous modulation of light and heat input for a building facade may be to design smart louver systems using responsive materials. Using a thin cast PDMS array with channels through which an index matching fluid could be flown, or replaced by air for inducing scattering, Park et al. designed a dynamic daylight control system capable of variable light scattering and redirection of light~\cite{PARK201487}. The system was not autonomous, however, and the switching between empty channels and channels filled with index matching fluid requires a closed system with pumps and tubing. We believe that light- or heat-driven LCE hinges can be combined with passive facade elements such as louvers in a way that may render the smart facade fully autonomous. Schwartz and co-workers demonstrated a concept for glare management based on vertical louvers organized into pairs, where the blades are split in two, with one half offset along the normal of the blade face~\cite{schwartzLouver}. By rotating the pair such that the normal to the louver surface always points to the sun, direct sunlight is fully blocked, while indirect sunlight and good visibility can be ensured as the parallel rays of the sun align only with the rotated louver, and indirect light is reflected inside (\textbf{Figure~\ref{subfig:left}}). The glare reduction is beneficial not only in reducing IR light and thus counteracting heat gain, but also through human comfort in visual glare. Unfortunately this type of heliotropic (sun-following) rotation mechanism is uncommon, most likely because of the expense and difficulty in manufacturing mechanical systems that are either preprogrammed, or equipped with sensors -- both of which require electrical power and wired electromechanical solutions. Designing an LCE actuator-based system that responds to the highest intensity light direction could be a way forward.

A simple solution might be to use an active LCE hinge that bends in response to sun irradiation, counterbalanced by a passive spring that pulls back under low light conditions. More elegant may be to place the louvers between two horizontal platforms, suspending the entire system around an axle using a ballbearing, such that it can easily rotate in response to subtle mechanical stimuli. We visualize this concept in Figure~\ref{subfig:right}. The top platform could be of glass, and below it we could pattern a circular array of LCE actuators, capable of protruding upwards under light illumination or local heating. To ensure sufficient stimulus, convex lenses may be used to concentrate the solar radiation onto each LCE. By adding small radially aligned opaque walls between each LCE actuator, only those that are facing the sun get irradiated and thus actuated. The top of the louver system would have a disc almost in contact with the platform, with a soft notch in one point. This notch would be locked in place by that particular LCE that protrudes the furthest into the notch. By the design, this happens to be the LCE pointing towards the sun, thereby providing the heliotropic rotation. An appropriately balanced spiral spring could act as a reset system each day. A suitable solution for the circular LCE actuator array would be stacks of LCE sheets with \textbf{n}(\textbf{r}) patterned into concentric rings, which can displace loads greater than 2500 times their own weight a distance of 0.5~mm~\cite{Guin2018}.  To make the system less sensitive to gusts of wind, it may be of interest to suspend the axles in a strongly shear thickening liquid, such as the colloidal suspensions used for 'liquid armor'~\cite{EgresJr2004}, which would counteract rapid motion due to wind but allow the slow motion driven by the LCE.

With louvers orienting autonomously to follow the sun, a natural further extension is to cover them with PV cells to also use the facade for generating electricity. Rather than today's large and heavy PV panels that typically cover roof tops, a more useful option would be the miniaturized micro-PV cells used in arrays, often in combination with optical elements to concentrate the sunlight (Concentrator Photovoltaics, or CPV), which are a hot current area of research~\cite{Alves2019}. These can have a lateral extension as small as 0.2~mm~\cite{Siopa2020} and be very thin and light-weight, making it easy to pattern louvers with such PV cells. Most importantly, considering the scope of this article, each micro-PV cell could be attached to the louver via an LCE hinge that lifts the cell with its active side out during the day, while illuminated by light, but folds it down during night, when it cannot generate electricity anyway. The active side is then protected from snowfall, sand or dust, and should such debris have accumulated on the active side during the day, it may fall off by gravity as a result of the bending. This concept for autonomous regular PV cell cleaning could of course be extended to other contexts, like replacing roof-top solar panels with LCE-driven selfcleaning CPV systems, largely eliminating the need for the very dangerous action of cleaning solar panels on the roof of a home. It may also be desirable to realize it in an Actuation-on-Demand version, as the need for electrical power is not an issue here. In particular in large solar parks in desert areas it could be very powerful to fold all PV panels away ahead of an incoming sand storm. In this context, it is worthwhile mentioning that LCE-driven heliotropic solar cells of rather large scale have already been realized~\cite{Li2012b,Guo2021}, albeit of quite different designs and functionalities than what we here propose.

\subsubsection{Regulating ventilation in response to heat and humidity variations}

Of similar importance to occupants of a building, and to their comfort, is the role of wind and air circulation. Functional materials-based solutions have included the use of Metal-Organic Frameworks (MOFs) as adsorbers with exceptional water uptake capacity~\cite{QIN2020106581}, but such approaches do not exchange the air or remove the humidity; the MOF may still be saturated with water. Here we discuss the capabilities of LCEs as a dynamic solution to provide a true air exchange and water removal, without energy consumption. Tuning LCEs to actuate at a temperature above or below common room temperatures could facilitate dynamic facades that prioritize natural air and ventilation when possible (a particularly attractive feature for hygiene), while minimizing the energy usage of HVAC (Heating, Ventilation and Air Conditioning) systems. While it would be valuable to be able to construct entire exterior walls with this functionality, also more modular or localized components could have significant impact. Kinetic buildings could be pushed to a new standard that further breaks down the relationships between inside and outside through building skins that actuate when sensing humidity. Smaller applications such as awnings or coverings at bus stations could provide airflow and semi-open ceilings, and then close to provide coverage when humidity reaches a point of rain.

\begin{figure}[!htbp]
\centering
\includegraphics[page=2, width=.6\textwidth]{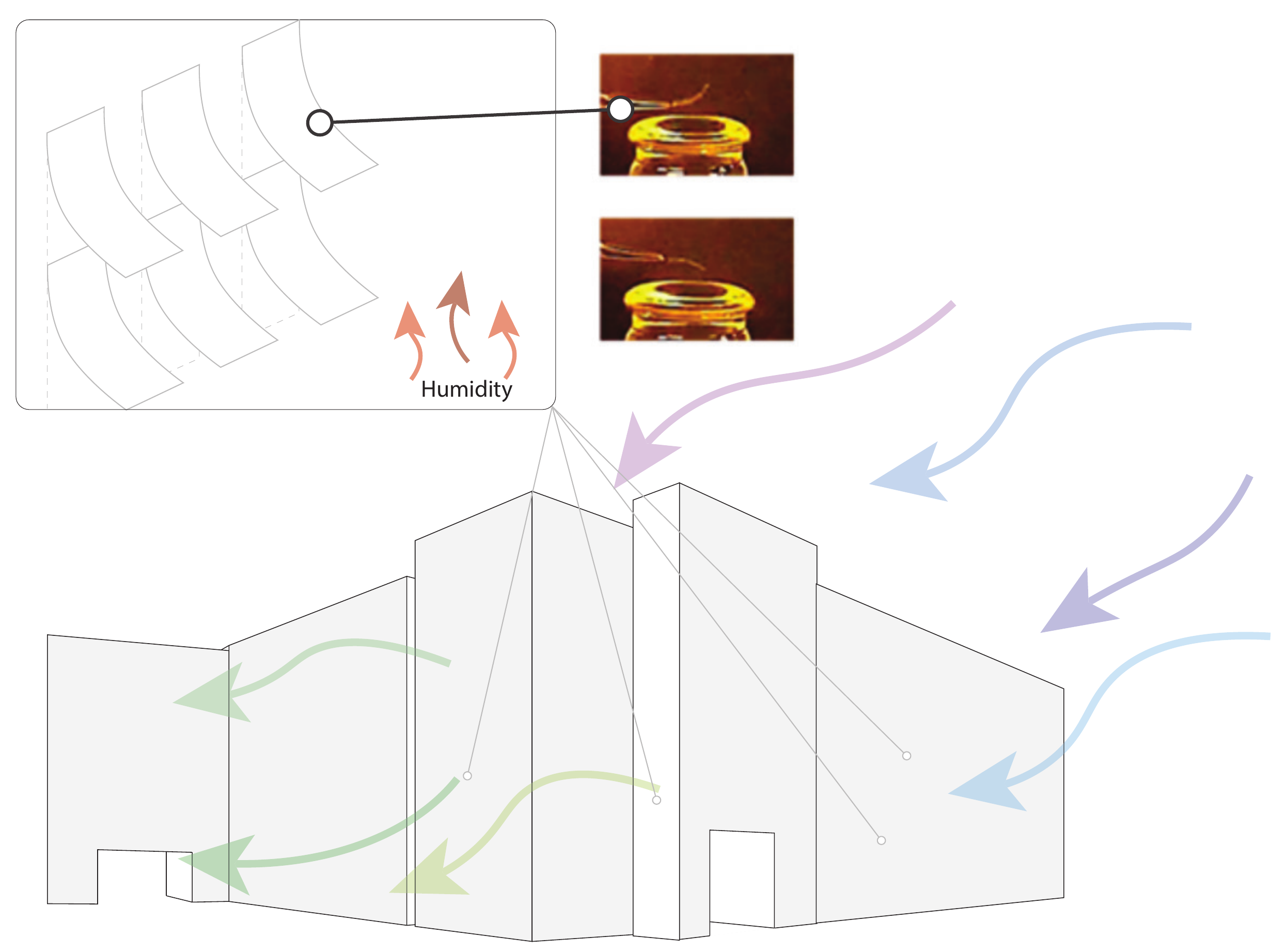}
\caption{ Concept of a facade opening as humidity increases. The envelop of a building is shown as a concept covered in LCEs that actuate as humidity increases, allowing air to circulate throughout. The photos of a humidity-sensitive LCE strip bending away from a vial filled with water when approaching the water surface (adapted with permission from \cite{deHaan2014}, copyright (2014) American Chemical Society) show an example of the technical implementation of the concept.}
\label{fig:airflowConcept} 
\end{figure}

In parts of the world where concrete and stone housing is common, even more interesting than responding to outdoor high humidity during rain may be to activate ventilation in response to \textit{indoor} high humidity, e.g. in kitchens and bathrooms. Mold formation inside living spaces is an enormous problem in many non-wooden houses during the cold part of the year. Insulated walls keeping the home warm also block the path for removing excess humidity, with the consequence that liquid water condenses from the air on cold surfaces, e.g. at corners of rooms with exterior walls. This forms an ideal environment for mold growth, with devastating health consequences, often related to allergies. In kitchens, the humid air arising during cooking fills fridges and freezers every time they are opened, their low temperature leading to significant water condensation or ice formation on the surfaces of the items in the fridge, which may become inedible as a consequence. Such food spoilage, as well as the increased electrical energy consumption of a fridge or freezer with ice-covered inner walls, could be avoided by dynamic ventilation, in the kitchen walls and possibly even in the concerned household appliances. 

Humidity-responsive LCEs might offer a way of realizing such dynamic ventilation (\textbf{Figure~\ref{fig:airflowConcept}}). The Schenning and Broer groups demonstrated LCE sheets that bend towards or away from humid air, depending on orientation~\cite{deHaan2014Humidity-responsive}. Many such LCE patches could be placed on a grid such that they form a closed wall when the air is dry, maximizing insulation and thus keeping heating/cooling costs low. However, if the internal humidity rises, the LCE sheets bend away from the humid air, opening channels for air exchange with the environment. While this obviously also reduces the thermal insulation efficiency, it is a temporary measure that autonomously stops as soon as the humidity has been reduced to acceptable levels. The channels opened during LCE actuation can be kept very small, hence air flow is enabled without allowing insects to enter. As neither energy nor user intervention is required to run the process, it is fully autonomous.

A particularly interesting example of water-responsive LCE is the humid air-driven cellulose-based motor developed by Geng et al~\cite{Geng2013}. The team prepared a ribbon of hydroxypropylcellulose (HPC) that bends away from humidity, and closed it into a loop. When one side of the loop is exposed to humid air, the local bending in conjunction with the topological constraints of the self-closing loop leads to a continuous motion of the ribbon. When suspended on two wheels, the ribbon rotates around the wheels as long as there is a gradient in humidity in the surrounding air. Not only can the humidity gradient thus, in principle, be used to do work but, perhaps more importantly, the HPC ribbon should be an ideal transporter of humidity out of a humid environment, as it autonomously moves humid HPC out to dry air and moves dry HPC into the humid environment, until the humidity gradient has been evened out. If such a ribbon in a ventilation shaft would be combined with the equally autonomous humidity-driven shutters describe above, a very efficient way of improving indoor air quality and saving energy overall might be realized.

\subsection{Sensing by Actuation}	
We end this section by discussing some very beneficial uses of LCEs 'operated in reverse', in the sense that an externally imposed mechanical strain changes the properties of the LCE, giving it the properties of a strain sensor. To this end we focus on CLCEs, the color of which can be tuned by strain. A compression along the helix, or an extension perpendicular to it, reduces $p$, thus blueshifting the color. This includes a blueshift from the IR region, i.e. the strain leads to the appearance of color in a CLCE that was colorless transparent in the relaxed state. Significantly, the response is local with negligible granularity, hence it can be used for strain mapping with very high resolution~\cite{Martinez2020Reconfigurable,Kizhakidathazhath2020Facile}. To illustrate the usefulness of CLCE-based strain mapping, we consider applications in safety, the existence of which could save many lives in accidents or natural disasters. We see potential in this area for the incorporation of CLCEs that change color under strain. They could be used as an autonomous system giving a clearly visible warning at critical areas, directly alerting the occupants of an impending danger.

\subsubsection{CLCE coatings alerting building occupants of a pending collapse of a roof, false floor or window}

The weight of snow on a roof can be very significant in cold parts of the world and sadly, accidents where roofs cave in, sometimes killing people underneath, are not as uncommon as one would wish. A notorious examples is the Katowice Trade Hall roof collapse in 2006~\cite{katowice}, leaving 65 people dead and more than 170 injured. Although there are many building standards that exist, they are largely a means to defining a minimal risk threshold and are by no means all encompassing. For example the northeastern United States had numerous building failures during a snow storm in 2011, and a lack of national safety guidelines encouraged a report on snow loads by FEMA~\cite{femasnow}. 

At least in case of glass roofs, it would be relatively easy to coat the glass plates with a CLCE film tuned for infrared selective reflection in the relaxed state, leaving it transparent in the absence of mechanical strain. In contrast, with the weight of snow on the film, $p$ would be compressed and the reflection blue-shifted. With properly tailored CLCE properties, the glass roof would start to reflect red when the weight on the roof is at the limit of what it can safely withstand, alerting occupants of an unexpectedly heavy load. The ability of CLCEs to present as a transparent material until actuated, provides unique opportunities, combining aesthetics with potentially life-saving functionality. The CLCE would provide crucial information during a time-sensitive situation, making the occupants aware that they must evacuate the building. 

Note that the CLCE would normally be applied on the outside of the glass, as this leads to significant compression of the soft CLCE as it is confined between the hard glass and the external load. The relative change in CLCE reflection wavelength is equal to the compressive strain \cite{Kizhakidathazhath2020Facile,Geng2022}, hence a small but detectable color shift from red at $\lambda=620$~nm to orange at $\lambda=590$~nm, comprising a wavelength reduction of about 5\%, requires the same amount of compression strain. While the bending of the glass pane under load might induce a color change of a CLCE applied to the glass inside, the response to bending is more complex and it is far from certain that the CLC experiences sufficient strain while the load on the glass is still within the limits of safe operation.    

\begin{figure}[!htbp]
\centering
\includegraphics[page=4, width=.8\textwidth]{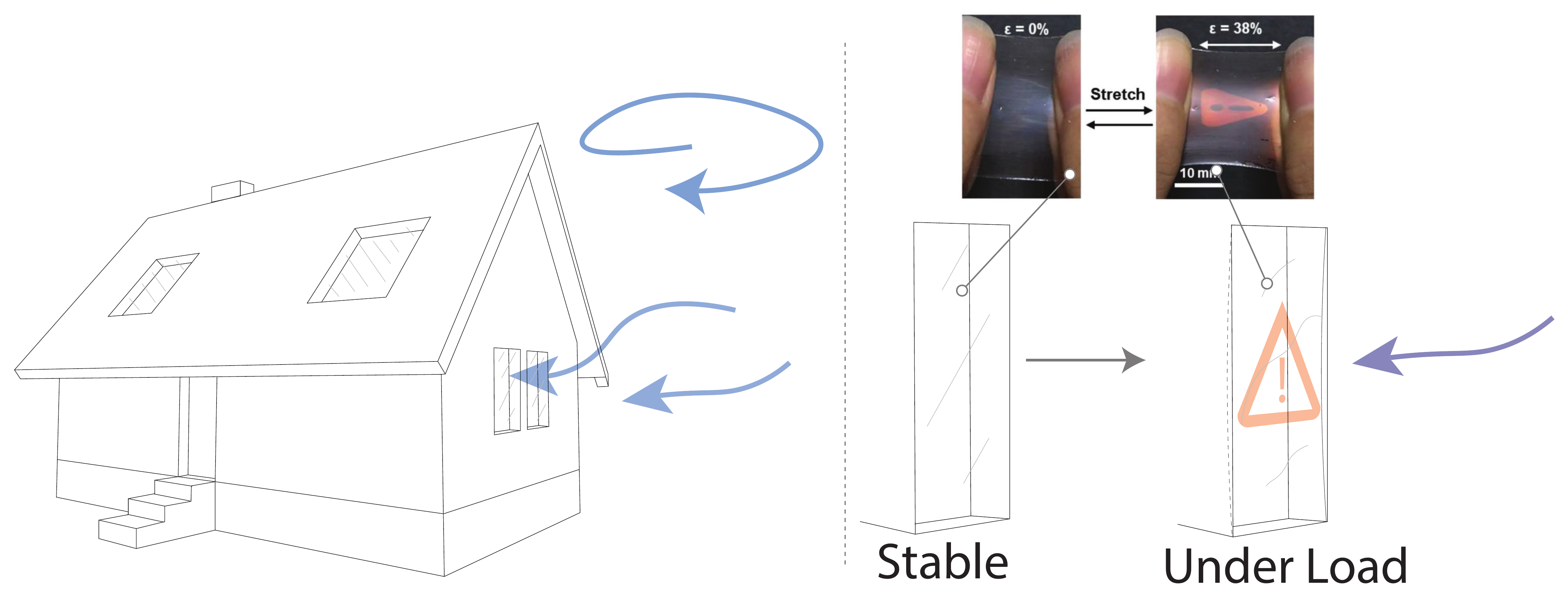}
\caption{ Concept of a CLCE coated window that flashes a warning when strong winds deform it to a critical level. The ability to show a pattern or image during strain is demonstrated in the two photos, reprinted with permission from~\cite{Zhang2020}, copyright (2019) Wiley-VCH.}
\label{fig:glasswanringConcept} 
\end{figure}

Along these same lines, CLCE coatings could be applied to the outsides of all types of glass panels. In areas of the world where hurricanes or tornados are a risk, having CLCE-coated storm windows would give visual indications of how much force was being applied (\textbf{Figure~\ref{fig:glasswanringConcept}}). They could possibly be tuned to provide a threshold in which there is danger of the window breaking, or even a warning threshold for when to seek underground shelter. To protect the soft CLCE from damage, it would ideally be sandwiched between the glass and a hard coating of durable transparent plastic, for instance polycarbonate.

To apply this approach to other roof styles such as concrete slabs, the CLCE must obviously be mounted on the inside. Here it would not be used to visualize the compressive force arising from the external load, but it could instead serve to visualize crack formation at the earliest stage of roof failure. Even the smallest cracks would lead to very strong local tensile strain of a CLCE coating, yielding a color change that allows occupants to see the cracks.

Even more informative than a general color change would be to pattern warning signs into the film, as demonstrated by de Hahn et al.~\cite{Zhang2020}, reproduced in Figure~\ref{fig:glasswanringConcept}. Here a CLCE film with its helix axis perpendicular to the film plane was designed to be featureless in the relaxed state, but upon stretching perpendicular to the helix near to its limiting extensability, making rupture a risk, a red warning triangle appeared, alerting the user that the tensile strain is now on a dangerous level. As demonstrated by Kizhakidathazhath et al.~\cite{Kizhakidathazhath2020Facile}, tensile strain perpendicular to the helix elicits the same mechanochromic response from CLCEs as compressive strain along the helix, hence the same patterning principle could be used for glass roofs. At the same time, there are many situations where rupture of an elastic component under tensile strain can have catastrophic consequences, and where human observers would notice a color change, hence this functionality could be applied in many contexts.

In the same way the strain in construction materials themselves could be indicated by means of CLCE films coating critical elements. Again, they could be designed for colorless transparency under normal conditions, but in case of excessive strain, a color would appear. Since the color response is nearly immediate and fully reversible~\cite{Kizhakidathazhath2020Facile,Geng2022}, the CLCE film would be well adapted to visualize dangerous dynamic loads. The CLCE could, for instance, be applied to supporting elements of false-floors, which are typically used as stages or audience platforms in concert halls or convention centers, the users often unaware of the designated maximum capacity. Again such instances can lead to tragic accidents, an example being the 2012 collapse of an audience platform in the Stockholm Globe Arena in Sweden, not withstanding the synchronized jumping during a concert by the artist Avicii. If the platform would have been coated with CLCE sheets generating reflections of different colors, security staff would have been able to see that the load was reaching dangerous levels before the accident occurred, allowing them to take evasive action.

\subsubsection{CLCE coatings mapping out the deformation of buildings during and after an earthquake }
In many parts of the world, seismic activity has proven to cause catastrophic building failures, resulting in numerous lost lives. During this activity, extreme deformations of the building could cause CLCE-coated elements to flash in a specific color, or even with a predetermined message. This type of user-environment interaction, through embedded intelligence of materials responding through the building's actuation, could aid in emergency route planning as signage dynamically appears based on the specific conditions of that emergency, guiding occupants to the safest places or, if possible, the exits. 

Perhaps even more valuable would be the post-earthquake situation, where CLCE coatings would greatly facilitate the evaluation of the building integrity. The common visual inspection done by inspectors and engineers could be greatly sped-up and improved by clear indicators that are now displaying warning colors/signs that show---even to a high degree of accuracy---remaining deformations and changes of the building compared to the original state. Significantly, for such an application the films could also be mounted in areas like service shafts, where also the regular pre-earthquake state does not need to be transparent. This maximizes the flexibility in tuning the color gamut of the response as well as the relaxed characteristics. Indeed, having the films colored from the beginning would facilitate the work of the inspection teams, as they will easily find them, whether the building has been deformed or not.

\section{Outlook}

We present a few of the many ways LCEs can be used in substantive and creative applications for improving the environments in which we live and work. Along with these various applications come the challenges and research needed to fully explore the potential. From a building perspective; cost of materials, both in the sourcing but also in the integration and installation, is critically important.  There is also a need for Architects and Designers to better understand how materials perform and the bounds of this performance. Currently, licensed Architects and Structural Engineers learn about the properties of typical building materials such as concrete, wood, and steel, but from a safety and reliability standpoint, LCEs (and for that matter, most shape memory polymers) are currently out of sight.

The size and longevity of buildings is also an issue not often discussed in the LCE community, given the focus primarily on robotics, with much shorter life cycle. Considering that many of these applications assume permanent or at least semi-permanent installations over decades, we must understand how the very same natural elements that we here consider the actuation triggers for LCEs might also cause failures or erode the polymers' ability to actuate and sense. It is not always possible to apply the LCEs on the interior of buildings, for instance, on building envelopes that are dynamically adjusting to exterior humidity and temperature variations for modulating air flow. Long-term exposure to sunlight, large variations in temperature, strong winds, rain, snow and even hail are conditions that the LCEs, at least up to a point, must sustain when applied on a building exterior. From this point of view, as well as from the point of view of minimizing the financial and ecological costs of massive scale-up of LCE production, it is noteworthy that bioderived LCEs have started appearing on the research arena. The humidity-driven motor ribbon of Geng et al.~\cite{Geng2013} was made of HPC, the same cellulose derivative that is commonly used as excipient in the pharmaceutical industry. Very recently, MacLachlan and co-workers succeeded in making the first mechanochromic CLCE based on cellulose nanocrystals~\cite{Boott2020}. Although this was still a hybrid of oil-based and cellulose-derived components, such developments hint at future LCEs which can be sustainably produced and which are less sensitive to weather, given that cellulose is the main component of all plants on Earth. Cellulose-based liquid crystals stabilized by soft water-soluble polymers have also been demonstrated as humidity-responsive actuators \cite{Wu2016}, even if they are not classical LCEs. With the current strong interest in liquid crystals based on cellulose, chitin or other polysaccharides \cite{Almeida2018,Parker2017}, it is likely that we will see this class of materials being used more and more to produce LCEs in a sustainable manner and at low cost.

Before LCEs can be realistically considered for many of the scenarios sketched in this article, several of these practical issues must be addressed by LCE materials scientists and engineers. While many recent developments are still far from mature, basic functionalities such as linearly contracting/expanding LCE strips and fibers capable of lifting loads, and sheets bending as active hinges have now been around for about two decades. We believe the time is ripe for an active effort to bring these types of LCE devices to market, thus taking on the chemistry and engineering hurdles on the way to mass production at low cost and with high reliability. While the solutions of these problems may not involve the fundamental research challenges that typically lead to high-profile scientific publications, they are absolutely essential for the successful practical implementation of LCEs in the built environment, or elsewhere for that matter. We also hope that industry will engage, leveraging the highly mature state of the most basic, and very useful, LCE actuators and sensors in order to add LCE hinges and pullers/pushers to their product catalogs. When architects can design with LCE devices in mind and engineers can build with them, their true potential can be unveiled. 

During a recent workshop among European researchers working on LCE actuators, a discussion about the transition from the lab to large-scale applications concluded with the insight that this transition is not blocked by any fundamental limitation, but rather by a 'chicken and egg' problem. All components for making LCEs could be made at rather low cost if industry would invest in large-scale production, but no company is ready to do so until a clear customer base exists. Meanwhile, architects and designers aware of the promise of LCEs would be ready to design for LCE integration, but as long as they cannot be certain that the LCEs they need are commercially available, they cannot take this step in actual projects. A viable strategy appears to be to focus first on thin-film LCEs, in the shape of 1D fibers, 2D sheets or empty shell structures as 3D examples, since these require a minimum amount of material and are thus the most likely to be manufacturable at acceptable cost. If they break through on the market, this can motivate industry to scale up materials production, bringing down the costs to the point where also bulk LCEs can be considered.

With the numerous research papers and experiments elaborating on the ways of developing LCEs, mass adoption can now be the next immediate focus. Part of this process includes a standardized description of material characteristics and performance: an MSDS for LCEs. How long does a certain LCE last in the sun, at different temperatures, with certain binders, connected to metal or plastic or wood?  Creating a catalog with systematic expression of material characteristics---how much does it deform in response to what stimulus? how soft is it in each state? etc.---would allow architects, designers, and general populace to easily connect the material itself to the creative and unique applications that propel society into the future. 

If the technical challenges and expected costs seem prohibitive at present, we should remember how an earlier liquid crystal-based technology, that of displays (LCDs), developed. The LCD was first introduced in 1964, with the only commercial product by 1973 being a pocket calculator screen. It was not until 1988 that a full-color and motion-capable display was created that truly launched LCD technology~\cite{kawamoto2012history}. Still, in the early 1990s few people thought LCDs would ever compete realistically with cathode ray tube displays and when large-area color LCDs eventually appeared, they had costs relegating them to the realm of luxury goods. But as technical challenges initially deemed insurmountable were solved, and chemistry was tailored for high-speed, high-contrast operation across an enormous temperature range, LCDs became the all-dominating display technology, with prices coming down to the lowest grade of consumer technology. From 2001 to 2011, the market value of LCDs grew from 20 to 120 billion US dollars~\cite{kawamoto2012history}. It is thanks to this development that the IT revolution became possible, as LCDs enabled small, high-resolution information output devices anywhere and everywhere. What future revolutions will LCEs enable, when the key actors have joint forces and brought LCEs to a standard, reliable and readily available technology?

\textbf{Acknowledgements} \par 
JPFL gratefully acknowledges financial support from the European Research Council under the European Union Seventh Framework Programme (FP/2007-2013)/ERC Grant Agreement no. 648763 (consolidator project INTERACT) and the Horizon Europe Framework Programme/ERC Grant Agreement no. 101069416 (Proof of Concept project REVEAL).

\bibliographystyle{unsrt}  
\bibliography{references}

\end{document}